\documentclass[vecphys]{svmult}

\usepackage{makeidx}         
\usepackage{graphicx}        
\usepackage{multicol}        
\usepackage[bottom]{footmisc}

\makeindex             

\def\rr{ {\vec r} }
\def\RR{ {\bf R} }

\def\FF{ {\bf F} }

\def\FF{ {\bf F} }

\def\uu{ {\bf u} }

\def\hh{ {\bf h} }
\def\XX{ {\bf X} }
\def\xx{ {\bf x} }
\def\mh{ {\underline{h}}}
\def\meta{{\underline{\epsilon}} }
\def\mt{ {\underline{t}}\  }
\def\mht{ {\underline{h}^T} }

\def\mepsilon{ {\underline{\epsilon}}\ }
\def\msig{ {\underline{\sigma}}\  }
\def\mSigma{ {\underline{\Sigma}}\  }
\def\mG{ {\underline{G}} }
\def\m1{ {\underline{1}} }

\def\mT{ {\underline{T}} }
\def\mC{ {\underline{C}} }
\def\rhohat{ {\hat{\rho}} }
\def\br{{\bf r}}

\begin{document}

\title*{Microscopic elasticity  of complex systems}
\author{Jean-Louis Barrat\inst{1}}
 \institute{Laboratoire de Physique de la Mati\`ere Condens\'ee et
 Nanostructures.
 Universit\'e Claude Bernard Lyon I and CNRS, 6 rue Amp\`ere, 69622 Villeurbanne Cedex,
 France \texttt{barrat@lpmcn.univ-lyon1.fr}
}
\maketitle

\section{Introduction}

Elasticity has the reputation of being a rather boring and
un-physical subject. In fact, dealing with second  (stress-strain)
or fourth (elastic constants) rank tensor guarantees that the
notations are in general rather heavy, and that the underlying
physics is not easily captured\footnote{To appreciate the very
nice physical content of elementary elasticity theory, the reader
is referred to the "Feynman lectures on Physics" or in a more
formal style  to the first chapters of the Landau and Lifschitz
textbook\cite{landau}}. The elastic stress-strain behavior is,
however, a very basic property of all solid materials, and one
that is rather easy to obtain experimentally. Hence simulation
methods for solid systems (hard or soft) should in general
consider the obtention of these elastic properties. The
determination of the response of a system to an imposed external
stress is an important topic, which was first addressed in the
simulation community through the pioneering work of Parrinello and
Rahman \cite{Parrinello81,Ray84}. While the main issue addressed
by the Parrinello-Ray-Rahman method was that of allotropic
transformations in crystalline solids, the current interest in
nanostructured materials opens a number of new questions. What are
the appropriate measures of stress and strains at small scale? Can
one scale down the constitutive laws of macroscopic elasticity,
and if yes to what scale ? Are the elastic constants of nanometric
solids identical to those of the bulk? What about the
elastic/plastic transition at small scales ? Many of these
questions are still unanswered, and the object of active studies.
In the following, I will describe recent studies that investigate
some of these questions.  After briefly recalling the appropriate
definitions, I will particularly concentrate on the case of
amorphous systems, in which some surprises arise due to the
non-affine character of deformations at small scales. In the last
section, I will describe how the Parrinello-Rahman scheme can be
extended to systems in which the particle representation has been
replaced by a field representation, as is common in polymer or
amphiphilic systems that self organize at a mesoscopic scale.

\section{Some definitions}

A convenient way to study elasticity with a microscopic viewpoint
is to think of  systems contained in periodic cells with variable
shape. This approach, which was initiated by Parrinello and
Rahman, is useful from  a practical, but also conceptual
viewpoint. In a system with periodic boundary conditions, the
simulation cell can be defined by three (in general nonorthogonal)
independent vectors $\hh_1$,$\hh_2$,$\hh_3$ forming the sides of
the parallelepiped cell. The Cartesian coordinates of these
vectors can be used to construct  a $3\times3$ matrix $\mh$
defined by $\mh = (\hh_1 , \hh_2 , \hh_3 )$. The Cartesian
coordinates of any point $\RR$ in the cell can be expressed as
\begin{equation} \label{rx}
\RR = \mh \XX
\end{equation}

 where $\XX$ is a rescaled vector whose components
lie in $[0,1]$. Integrals on $\RR$ can be converted into integrals
over $\XX$ by using a scaling factor $\det \mh$, which represents
the volume of the cell, $V$. In the case of a particle or monomer
number density, for example, one can write
\begin{equation}
\rho(\RR)= \rho(\XX) ( \det \mh)^{-1}
\end{equation}
The metric tensor $\mG$ is constructed from $\mh$ as
\begin{equation}
\mG = \mht \mh
\end{equation}
where $\mht$ is the transpose of $\mh$.  $\mG$ is used in
transforming dot products from the original Cartesian to rescaled
coordinates, according to
\begin{equation}
\RR\cdot \RR^\prime = \XX \cdot \mG \cdot \XX^\prime = X_\alpha
G_{\alpha \beta} X^\prime_\beta
\end{equation}
where here and in the following summation over repeated indexes is
implicit.

Elasticity theory describes the deformation of any configuration
from a reference configuration in terms of a strain tensor. This
tensor is constructed by relating the vector connecting two points
in the deformed configuration to the corresponding displacement of
the same points in the reference configuration. If the reference
configuration of the simulation box is denoted by $\mh_0$,  the
displacement is $\uu=\RR-\RR_0= (\mh\mh_0^{-1}-1)\RR_0$, and  the
strain is given by cite{Parrinello81,Ray84}
\begin{equation}\label{strain}
\meta = \frac{1}{2}\left[ (\mht_0)^{-1} \mht  \mh (\mh_0)^{-1}
-\m1 \right] = \frac{1}{2}\left[ (\mht_0)^{-1} \mG (\mh_0)^{-1}
-\m1 \right]
\end{equation}
where $\m1$ denotes the unit tensor. It is important to note that
this expression, usually known as the \textit{Lagrangian strain
tensor} is not limited to small deformations \cite{landau}.
Usually, the reference configuration $\mh_0$ will be defined as a
state of the system under zero applied external stress. If one
starts with a cubic cell, $\mh_0$ is the identity matrix and the
relation between $\meta$ and $\mG$ simplifies. The thermodynamic
variable conjugate to this strain tensor, in the sense that the
elementary work done on the system can be written in the form
\begin{equation}\label{dw}
\delta W = V_0 \mathrm{Tr} (\mt \delta \meta),
\end{equation}
is the thermodynamic \emph{tension} tensor $\mt$ \cite{wallace70},
also known as Piola-Kirchhoff second stress tensor. $V_0 \equiv
\det \mh_0$ denotes the volume of the system in the reference
configuration. This thermodynamic tension tensor can be related to
the more usual \emph{Cauchy stress tensor} $\msig$ through
\begin{equation}\label{tsigma}
    \msig = \frac{V_0}{V}   \mh \: (\mh_0 )^{-1} \mt (\mht_0 )^{-1}\mht
\end{equation}
The tension is the derivative of the free energy with respect to
the strain, which is calculated from the reference configuration.
The Cauchy stress, on the other hand, is the derivative of the
free energy with respect to an incremental strain taken with
respect to the actual configuration. This Cauchy stress tensor is
the one that enters momentum conservation and whose expression is
given by the usual Irving-Kirkwood formula for pairwise additive
potentials (see below).  The difference between these two
quantities can be understood, qualitatively, from the fact that
the strain is not an additive quantity, as can be seen from the
existence of the nonlinear term in equation \ref{strain}. While
the Cauchy stress has a mechanical meaning in terms of forces
within the sample, the thermodynamic tension is a purely
thermodynamic quantity, and does not in general have a simple
mechanical interpretation.

Fortunately, in the limit of small deformations which I will
concentrate on, the differences between these various expressions
of stress tensors can be forgotten. This is not the case, however,
for large deformations, where these differences result in a whole
variety of stress/strain relations and associated elastic
constants. This is especially important when dealing with solids
under high pressure, where for example one has to be careful as to
which of these elastic constants is used to compute e.g. sound
velocities. I will refer the interested reader to the reference
publication of Klein and Baron \cite{Klein} for an in depth
discussion of these subtleties.

\section{Finite temperature elastic constants: Born and
fluctuation terms}

The elastic constants for a material made of particles interacting
through a pair potential $\phi(r)$ (I'll keep this simplifying
assumption in the following) can be determined from simulations
using an approach presented by Hoover and coworkers \cite{Hoover}.
 These authors start from the explicit expression of the free energy in
terms of a configuration integral
\begin{equation}
\exp(-\beta F) = V^N \int d\XX_1 d\XX_2..d\XX_N \exp(-\beta
H(\{\mh \XX_i\}))
\end{equation}
where $H(\{\RR_i\})= \sum_{ij} \phi(\RR_i-\RR_j)$ is the total
interaction energy.

The derivative with respect to strain is taken using
\begin{equation}
dF = \mathrm{Tr} (\frac{\partial F }{\partial \mG} d\mG) =
2\mathrm{Tr }\left(\mh_0 \frac{\partial F }{\partial \mG} \mht_0
d\mepsilon\right)
\end{equation}
which gives for the thermodynamic tension matrix
\begin{equation}
V_0 t_{\alpha\beta} =N k_BT  \mh_{0,\alpha\gamma}
\mG^{-1}_{\gamma\delta} \mht_{0,\delta\beta} +
\mh_{0,\alpha\gamma}\langle \sum_{ij} X_{ij,\gamma}X_{ij,\delta}
\frac{\phi'(R_{ij})}{R_{ij}} \rangle
 \rangle \mh_{0,\delta\beta}
  = N k_BT  \mh_{0,\alpha\gamma} \mG^{-1}_{\gamma\delta} \mht_{0,\delta\beta}
  + \langle \hat{\mT}_{\alpha\beta}
 \rangle
\end{equation}
where $X_{ij}=X_i-X_j$, $\sum_{ij}$ is the summation over all
distinct pairs of particles, and the pair potential $\phi$ is
assumed to depend only on the particle separation $R_{ij}$. The
brackets $\langle\rangle$ denote a thermal average. This obviously
reduces to the usual Kirkwood formula for small deformations
($\mh_0=\mh$). Note that the first term arises here from the
volume factor $V^N$ in the configuration integral, but could also
be obtained by introducing the momenta and the kinetic energy
contribution in the partition function. The last term defines the
potential energy contribution to the microscopic stress tensor,
denoted by $\hat{\mT}$. Carrying out one more derivation with
respect to strain, one obtains the elastic constants in the limit
of zero strain (more general expressions for arbitrary strain can
be found in refs \cite{Lutsko,Maloney}):
\begin{equation}
 C_{\alpha\beta\gamma\delta} = \frac{\partial
t_{\alpha\beta}}{\partial \epsilon_{\gamma\delta}} = 2Nk_BT
(\delta_{\alpha\gamma} \delta_{\beta\delta}+ \delta_{\alpha\delta}
\delta_{\beta\gamma}) - \frac{V_0}{k_BT}\left[\langle
\hat{T}_{\alpha\beta} \hat{T}_{\gamma\delta}\rangle - \langle
\hat{T}_{\alpha\beta}\rangle  \langle
\hat{T}_{\gamma\delta}\rangle\right] +
C_{\alpha\beta\gamma\delta}^{Born} \label{cel1}
\end{equation}
Where the last (Born) term is written in terms of potential energy
functions
\begin{equation}
 C_{\alpha\beta\gamma\delta}^{Born} = {1\over
V_0}\langle\sum_{ij} R_{ij,\alpha}
R_{ij,\beta}R_{ij,\gamma}R_{ij,\delta}\left(
\frac{\phi"(R_{ij})}{R_{ij}^2}-
\frac{\phi^\prime(R_{ij})}{R_{ij}^3} \right) \rangle
\end{equation}
with $\RR_{ij}=\RR_i-\RR_j$. The term in square brackets in
\ref{cel1} is called the fluctuation term, and is generally
expected to be a correction to the main Born term at finite
temperature. We will see below that this term, even in the low
temperature limit, remains an essential contribution to the
elastic properties of disordered systems. Formulae  that
generalize the equations above to local stresses and elastic
constants can be found in references \cite{goldenberg,depablo}.

 Note that a different approach must be used for hard
core potentials. In that case, one possibility is to study strain
fluctuations either in a cell of variable cell \cite{nielaba}, or
to directly study the stress-strain relation in a deformed cell
\cite{frenkel}.

\section{Amorphous systems at zero temperature: nonaffine
deformation} The simulation of amorphous systems at low
temperatures  is interesting from the point of view of the physics
of glasses, but also because these models can serve as very
elementary examples of "athermal" systems such as granular piles
or foams. The elastic/plastic response of such complex systems is
not well understood, and is currently the subject of many
experimental studies \cite{clement,dennin,debregeas,graner}.

A naive approach to the calculation of elastic properties in such
systems would consist in taking the second derivative of the
potential energy $H=\sum_{i<j} \phi(R_{ij})$ with respect to
strain. Such an approach is easily shown to yield elastic
constants that correspond to  the Born expression, without the
thermal average brackets. Although such an approach seems natural
- the "fluctuation" term could be ignored at zero temperature -,
it proves in fact completely incorrect for disordered systems, or
even for crystals with a complex unit cell.

The essential point is that the derivatives have to be taken not
at constant $X$, but rather keeping the force on each atom equal
to zero in the deformed configuration. In other words, one has to
allow for relaxation of the deformed configuration before
computing the energy and stresses. It was shown by Lutsko
\cite{Lutsko} (see also the recent work by Lemaître and Maloney,
ref\cite{Maloney}) that this relaxation gives a contribution to
elasticity which is identical to the zero temperature limit of the
fluctuation term. The corresponding proof can be  briefly
summarized as follows.

The elastic constant is written as
\begin{equation}
C_{\alpha\beta\gamma\delta} = \frac{\partial
t_{\alpha\beta}}{\partial\epsilon_{\gamma\delta}}{\Large
|}_{\FF_i=0}
\end{equation}
$_FF_i=0$ indicates the constraint that forces on the particles
must remain zero during the deformation.  The variables in the
problem are the reduced coordinates, $\XX_i$, and  the strain
$\mepsilon$. The force $\FF_i$ is a function of these variables,
so that the constrained derivative above can be written as (for
simplicity, we drop in this formula and the following the Greek
indexes for Cartesian coordinates):
\begin{equation}
\mC =  \frac{\partial t}{\partial \epsilon}\vert_{\XX_i} -
\frac{\partial t}{\partial \XX_i}\left(\frac{\partial
\FF_j}{\partial \XX_i}\right)^{-1}\frac{\partial \FF_j}{\partial
\epsilon} \label{mC}
\end{equation}
where terms such as $\left(\frac{\partial \FF_j}{\partial
\XX_i}\right)$ have to be understood in a matrix sense. In fact,
$D_{ij}= \left(\frac{\partial \FF_j}{\partial \XX_i}\right)$ is
nothing but the \textit{dynamical matrix} of second derivatives of
the potential energy with respect to atomic positions. Finally,
using the definition of the force $\FF_j$ and of the tension $\mt$
as derivatives of the potential energy, equation \ref{mC} can be
rewritten in the more symmetric form
\begin{equation}
\mC =  \mC^{Born} - \frac{\partial t}{\partial
X_i}\left(\frac{\partial F_j}{\partial
X_i}\right)^{-1}\frac{\partial t}{\partial X_j} \label{mC1}
\end{equation}

The direct evaluation of the second term in this equation is not
straightforward, hence the actual procedure to obtain the zero
temperature elastic constants generally consists in carrying out
explicitly an affine deformation of all coordinates, then letting
the atomic positions relax (using e.g. conjugate gradient
minimization)\cite{NumRec} to  the nearest energy minimum.
Equation \ref{mC1} however can be used to show that the resulting
elastic constants are identical to those obtain at a finite, low
temperature using equation \ref{cel1}. The proof goes simply by
expanding both the stress and energy in terms of the atomic
displacements in the unstrained reference configuration
\begin{equation}
H= H_0 + \frac{1}{2} D_{ij} \delta X_i \delta X_j  \ \ ; \ \
 t= t_0 +
\frac{\partial t}{\partial X_i} \delta X_i
\end{equation}
Performing the resulting gaussian integrals, it is seen that the
"fluctuation" term in \ref{cel1} and the "relaxation" term in
\ref{mC1} are identical in the limit of zero temperature.

\section{Numerical results}
\begin{figure}[t]
\centerline{\resizebox{12cm}{!}{\includegraphics*{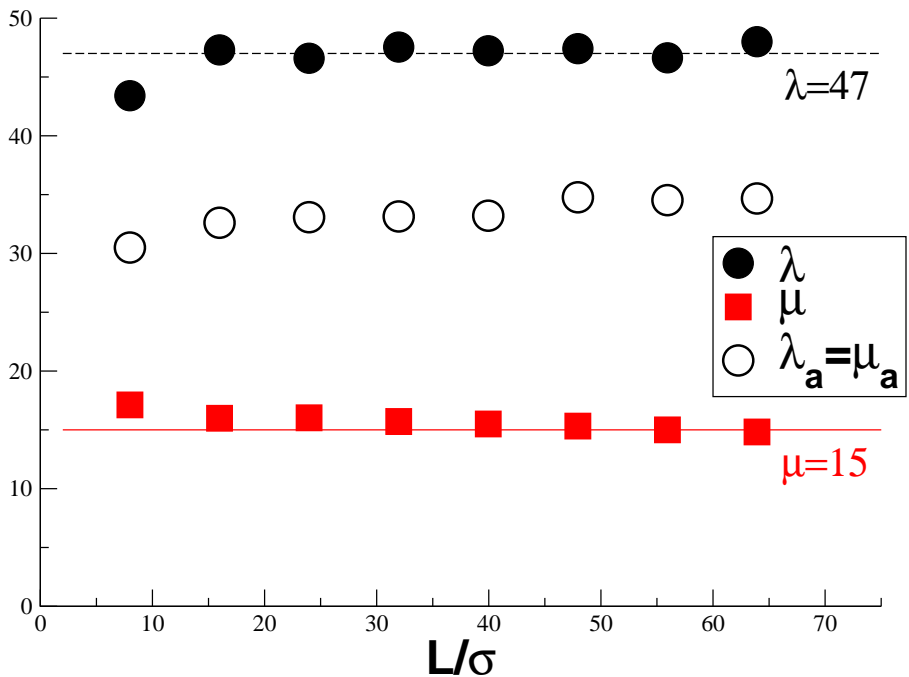}}}
\vspace*{-0.2cm} \caption[]{
Lam\'e coefficients $\lambda$ (spheres) and $\mu$ (squares) {\em
vs.} system size $L$, for a simple polydisperse Lennard-Jones
"glass". Full symbols correspond to a direct measurement using
Hooke's law with relaxation, open symbols correspond to the Born
approximation).
The effect of system size is weak. For large boxes we get $\mu
\approx  15$ and $\lambda \approx 47$.
 \label{fig1}}
\end{figure}

We now address the importance, qualitative and quantitative, of
this "relaxation-fluctuation" contribution. Quantitatively, the
importance of this contribution can be estimated from figure
\ref{fig1}. In this figure,  the Lam\'e coefficients of a three
dimensional, amorphous, Lennard-Jones system (see ref.
\cite{leonforte3}) at zero temperature are computed using the Born
approximation and the exact formula, equation \ref{cel1}. It is
seen that the relaxation term can account for as much as 50\% of
the absolute value of elastic constants. Although this fraction
may obviously be system dependent, the situation is very different
compared to simple crystals (with one atom per unit cell,e.g. FCC
in the Lennard-Jones system) in which the elastic constants are
exactly given by the Born term (see e.g. \cite{roux87} for a
comparison between amorphous systems and simple crystal
structures). The relaxation contribution tends to lower the shear
modulus $\mu$, and to increase the coefficient $\lambda$.
Remarkably, the bulk modulus $K=\lambda+2\mu/d \approx 57$ ($d=3$
is the dimensionality of space) would be correctly predicted by
the Born calculation.

\begin{figure}[t]
\centerline{\resizebox{12cm}{!}{\includegraphics*{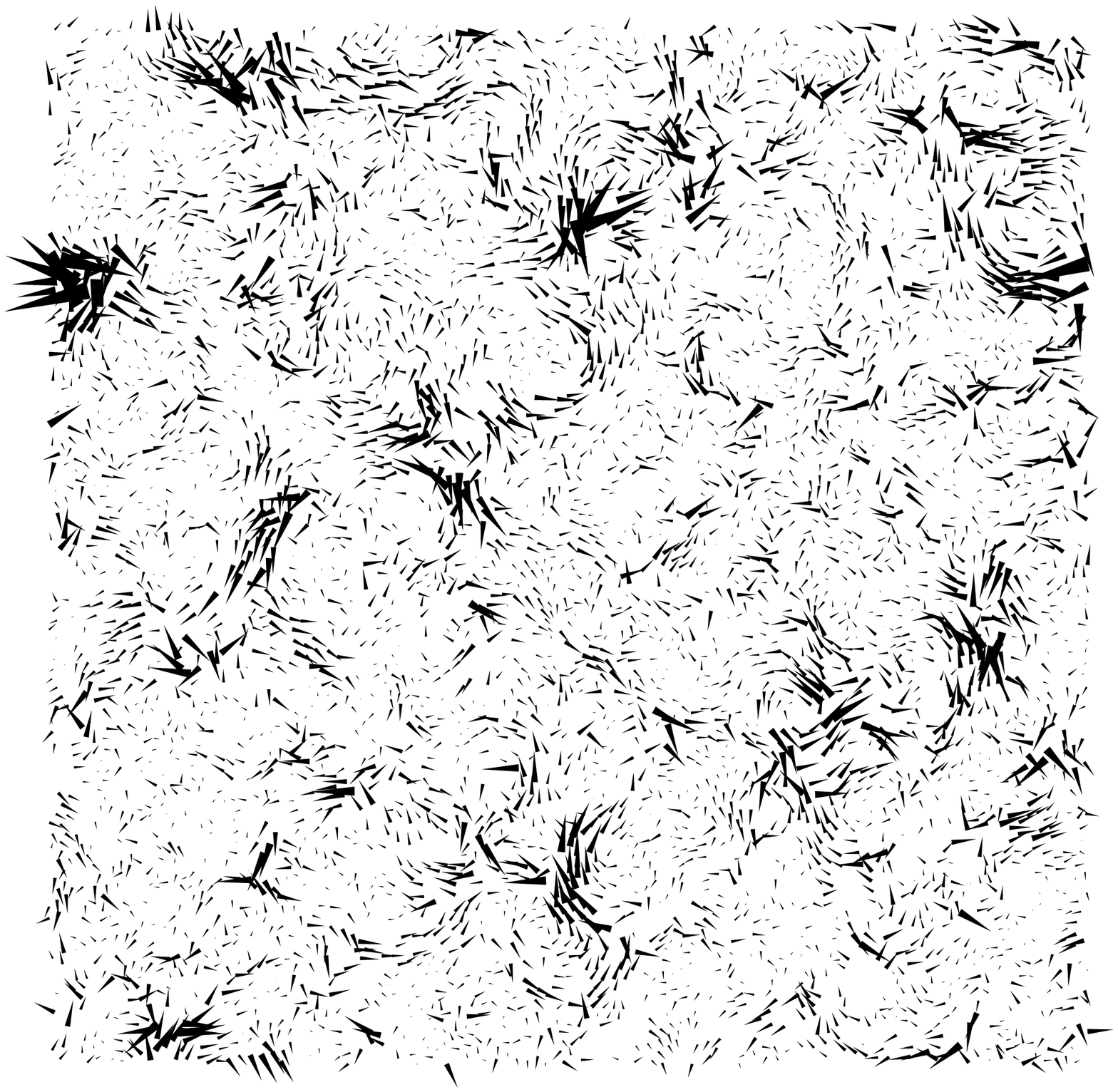}}}
\vspace*{-0.2cm} \caption[]{
Snapshot of the nonaffine displacement field in a 2d Lennard-Jones
amorphous system undergoing uniaxial extension. Note the large
scale, vortex like structures. The  sample contains about 20 000
particles.
 \label{fig2}}
\end{figure}

\begin{figure}[t]
\centerline{\resizebox{12cm}{!}{\includegraphics*{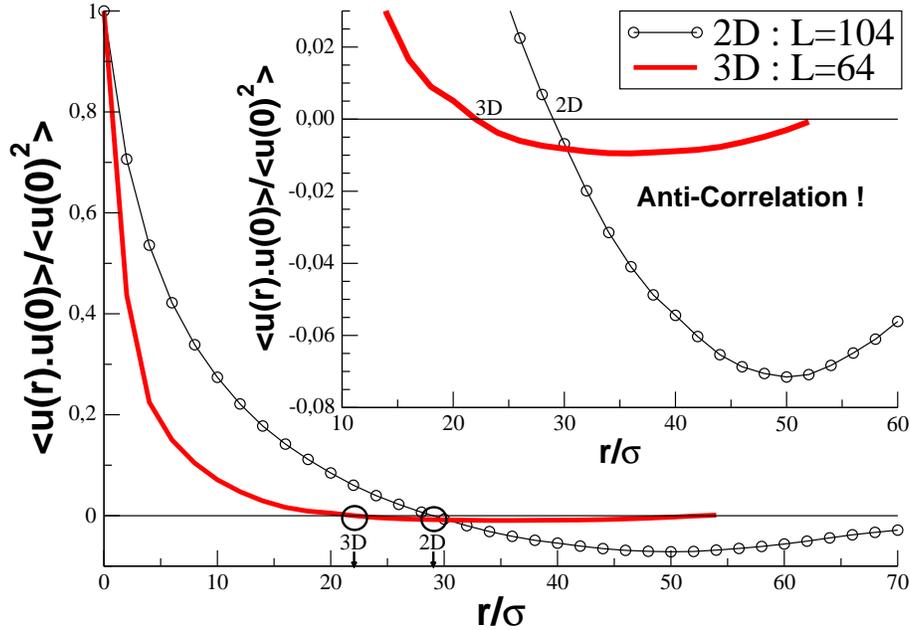}}}
\vspace*{-0.2cm} \caption[]{ Correlation function $C(r)=
<\uu^{NA}(\rr) \cdot \uu^{NA}(\rr)>$ of the nonaffine displacement
field $\uu^{NA}(\rr)$ in an amorphous sample undergoing a simple
uniaxial extension. Both the 2d and 3d case show correlations that
extend over scales  of typically 20-30 particle sizes. A negative
tail can be associated with "vortex like" structures that reflect
the essentially rotational character of the non affine field.
 \label{fig3}}
\end{figure}

As discussed above, the  Born formulae would be exact at zero
temperature if the global deformation was equivalent to an affine
deformation of atomic coordinates at all scales, i.e. a mere
rescaling of $\mh$ at fixed values of $\{X_i\}$. The failure of
the Born calculation can therefore be traced back to the existence
of a \textit{non-affine deformation field}, which stores part of
the elastic deformation energy. This field is defined by
substraction from the actual displacement of the atoms (after
relaxation) the displacement that would be obtained in the affine
hypothesis. The existence of this non-affine deformation field was
pointed out in several recent publications
\cite{goldenberg,langer,Liu,Wittmer}. In
\cite{Wittmer,leonforte04}, it was in particular shown that this
non-affine contribution is correlated over large distances, and is
organized in vortex like structures (i.e. is mostly rotational in
nature). These properties are illustrated in figures \ref{fig2}
and \ref{fig3}, for a simple Lennard-Jones two dimensional system.

A slightly different, more local and general, definition of the
nonaffine displacement field (or "displacement fluctuation") was
proposed in ref. \cite{goldenberg}. In this reference, the
nonaffine field is defined by substracting from the actual
displacement a local displacement field built that is obtained
using a coarse-graining procedure. This allows in principle to
deal with situations in which the displacement field has a complex
structure. In the case of simple shear considered here, our
definition should be sufficient. Reference  \cite{goldenberg} also
demonstrates that, even when the displacements are not locally
affine, there exists a local linear relation between the stress
and strain fields, at sufficiently large scales (resolution).
These fields are evaluated at the same position with a chosen
resolution. The derivation does assume that the displacement
fluctuations are uncorrelated over sufficiently large scales, i.e.
it will be valid at scales larger than the one discussed here.

 A quick study of a simple one dimensional  model is useful to
understand the importance of the nonaffine displacement field. Let
us consider a chain of $N$ atoms, connected by springs $k_i$,
submitted to a force $F$. The extension of spring $i$ (linking
site $i$ to $i+1$) is $\delta_i=F/k_i$. One can therefore write
the displacement of atom $p$
\begin{equation}
u_p= F \times \sum_{i=1}^{p} k_i^{-1}
\end{equation}
The affine displacement is just $u_p^{aff}= (p/N)\times F\times
\sum_{i=1}^N k_i^{-1}= p F <k^{-1}>$, where the $<>$ refer to an
average over the distribution of elastic constants and the large
$N$ limit has been taken to compute the affine displacement. As a
result we have for the nonaffine displacement of atom $p$
\begin{equation}
u_p^{NA}=u_p-u_p^{aff} = F \times \sum_{i=1}^P (k_i^{-1}
-<k^{-1}>)
\end{equation}
which shows that in this simple 1d situation the mean squared
value $<^(u_p^{NA})^2>\propto p <(\delta k^{-1})^2> $  is
increasing linearly with $p$, and proportional to the variance of
$1/k_i$ (see also the discussion by DiDonna and Lubensky,
cond-mat/0506456).

 The existence and
nature of the length scale over which the non affine field is
correlated is still  a matter of debate \footnote{In a recent
preprint (cond-mat/0506456, "Nonaffine correlations in Random
Elastic Media"), DiDonna and Lubensky argued that the nonaffine
field has logarithmic (in 2d) or $1/r$ (in 3d) correlations, and
hence no characteristic length scale. That such singular behaviour
is possible is already illustrated in the simple 1d example above.
Such a behaviour, however, is not evident in our numerical
results. Their calculation, based on the fact that elastic
propagator have $1/k^2$ behaviour in Fourier space, is
perturbative in the disorder strength, and it could be that we are
investigating a "strong disorder" limit.  In any case, further
investigations are needed to assess the actual existence of such
long range correlations}. Clearly, the correlation length $\xi$ is
a lower limit for the applicability of continuous elasticity
theory. This limit manifests itself in several different ways. If
one considers the vibrations of a system of size $L$, these
vibrations will be properly described by the classical elasticity
theory only if the corresponding wavelength is larger than $\xi$.
If one considers the response to a point force, this response will
be correctly described by the continuum theory only beyond the
length scale $\xi$. More precisely, it was found that the {\it
average} response is described by continuum theory essentially
down to atomic size, but that the {\it fluctuations} (from sample
to sample) around this average are dominant below $\xi$
\cite{leonforte04}.

\begin{figure}[t]
\centerline{\resizebox{12cm}{!}{\includegraphics*{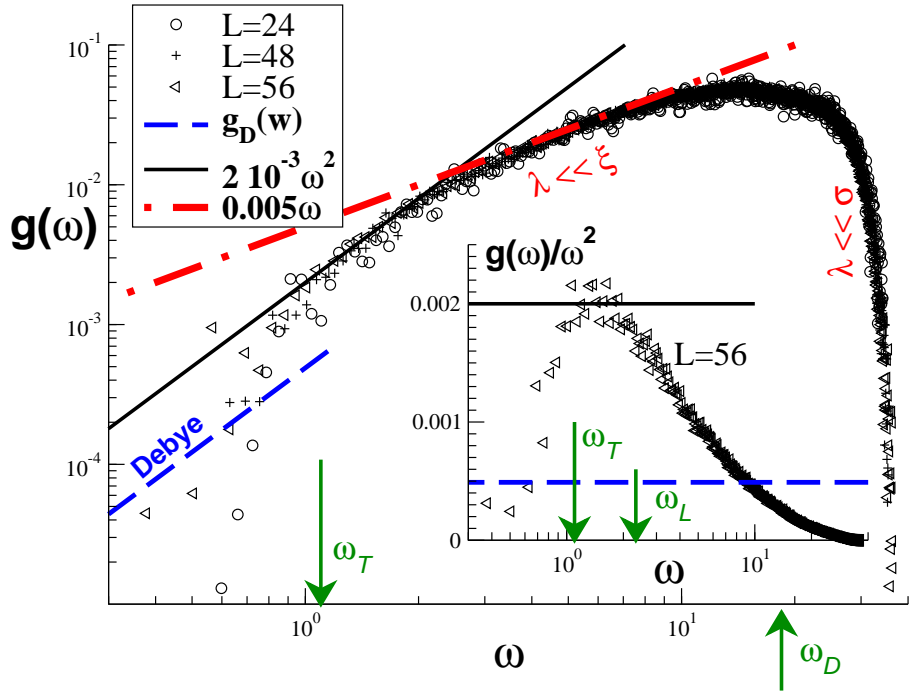}}}
\vspace*{-0.2cm} \caption[]{Vibrational density of states in a 3d
Lennard-Jones amorphous system. The "boson peak" is apparent as
the deviations from the Debye prediction in the ratio
$g(\omega)/\omega^2$. This peak is observed at frequencies of
order $2\pi c/\xi$ where $c$ is the speed of sound.}
 \label{fig4}
\end{figure}

Finally, it is very likely that the existence of this length scale
is related to a prominent feature of many disordered systems, the
so called 'boson peak'. This feature actually corresponds to an
excess (as compared to the standard Debye prediction,
$g(\omega)\propto \omega^{d-1}$ in $d$ dimensions) in the
vibrational density of states $g(\omega)$ of many amorphous
systems. This excess shows up as a peak in a plot of
$g(\omega)/\omega^{d-1}$ vs $\omega$, that usually lies in the THz
range. In terms of length scales, we found that this peak
typically corresponds to wavelengths of the order of magnitude of
$\xi$ (see figure \ref{fig4}). A simple description \cite{duval}
is therefore to assume that waves around this wavelength are
scattered by inhomogeneities, and see their frequencies shifted to
higher values. Pressure studies show that the boson peak is
shifted to higher frequencies under pressure, consistent with a
shift to smaller values for $\xi$ obtained in simulations. Another
very interesting evidence for the existence of mesoscale
inhomogeneities was recently provided by Masciovecchio and
coworkers \cite{ruocco}, by studying Brillouin spectra in the
ultraviolet range. The width of the Brillouin peak shows a marked
change for wavelength between 50 and 80 nm, indicative of
scattering by elastic inhomogeneities.

A very interesting question is whether this characteristic
correlation length for elastic inhomogeneities, which can reach
rather large values compared to atomic sizes, is somehow
associated with a 'critical' phenomenon.  An idea that was
recently suggested by Nagel and co-workers \cite{NagelWyart} is
that this correlation length should diverge at the so-called
'jamming' transition in purely athermal systems. The jamming
density is defined,  in a system with purely repulsive
interactions at zero temperature, by the density at which the
system will start to exhibit mechanical rigidity. Below the
jamming density  $\phi_c$, an infinitesimal temperature results in
diffusion, while systems above $\phi_c$ remain in a frozen state
on macroscopic time scales. Based on general arguments concerning
isostaticity of the packing at $\phi_c$, Nagel and coworkers
\cite{NagelWyart} suggested the existence of a correlation length
associated with soft modes, that diverges at the transition.
Although the arguments are in principle valid only for contact
type interactions, it would be quite interesting to follow the
evolution of $\xi$ for a system {\it with attractive interactions,
but under tension}, expecting perhaps a divergence close to the
rupture threshold.

Finally, let us mention that a different way of studying local
elastic properties was proposed by de Pablo and coworkers through
the study of local elastic moduli, which can be defined by using
the definition \ref{cel1} to a small,  finite box \cite{depablo}.
Depending on the scale at which they are measured, these moduli
can take negative values. Such regions would be unstable, if they
were not immersed in a matrix of "normal" regions. The size over
which local elastic constants are found to be negative is small
(typically 3 particle sizes), and could probably be considered as
a first coarse-graining scale for using classical methods for
disordered systems \cite{schirmacher}.

\section{Polymeric systems: stresses and self consistent field theory}

Self consistent field theory is a powerful approach to the
determination of phase equilibria in polymer systems with complex
architectures. The theory directly deals with density fields
rather than particles, and minimizes  a mean field like free
energy. The method is particularly suitable for polymers, in which
interactions on length scales comparable to the chain size are
effectively very soft. The method is well known and has been
described in many publications (see e.g. \cite{ghf02,matsen94})
and I will only describe briefly the main steps and the way a
constant stress method can be introduced in the simulation. This
option allows one to  obtain relaxed configurations at zero
imposed stress easily, or to study the effect of anisotropic
tension on phase behavior.

As a representative example, I consider a model for an
incompressible AB diblock copolymer melt \cite{matsen94}. The melt
consists of $n$ identical diblock copolymer chains composed of
monomer species $A$ and $B$ and is contained in a volume $V$. Each
of the chains has a total of $N$ statistical segments; a fraction
$f$ of these segments are type $A$ and constitute the A block of
each macromolecule. For simplicity, the volume occupied by each
segment, $v_0$, and the statistical segment length, $b$, are
assumed to be the same for the A and B type segments. The
Hamiltonian for this system can be written
\begin{equation}\label{h0}
H= \sum_{i=1}^{n} \frac{k_BT}{4 R_{g0}^2} \int_0^1 ds
\left(\frac{d\RR_i(s)}{ds}\right)^2 + v_0 \chi_{AB} k_BT \int d\br
\: \rhohat_A(\br) \rhohat_B(\br)
\end{equation}
where $\RR_i (s)$ with $s \in [0,1]$ is a space curve describing
the conformation of the $i$th copolymer and $R_{g0}^2 =b^2 N/6$ is
the radius of gyration of an ideal chain of $N$ statistical
segments. Interactions between dissimilar segments A and B are
described by the Flory parameter $\chi_{AB}$. The densities
$\rhohat_{A,B}(\br)$ are microscopic segment density fields
defined by
\begin{equation}\label{eq1}
    \rhohat_A (\br ) = N \sum_{i=1}^n \int_0^f ds \; \delta (\br -
    \RR_i (s) )
\end{equation}
and
\begin{equation}\label{eq2}
    \rhohat_B (\br ) = N \sum_{i=1}^n \int_f^1 ds \; \delta (\br -
    \RR_i (s) )
\end{equation}
A local incompressibility constraint $\rhohat_A (\br ) + \rhohat_B
(\br ) = \rho_0$ is imposed in this standard copolymer melt model
for all points $\br$ in the simulation domain. The total segment
density $\rho_0$ can evidently be expressed as $\rho_0 =nN/V =
1/v_0$. Using the rescaled coordinates  $\XX(s)$ (taken in
$[0,1]^3$),  the generalized partition function that has to be
sampled\textit{ for a fixed value of the thermodynamic tension}
$\mt$ reads
\begin{eqnarray}
 Z & = & \int d(\mh)  (\det \mh)^{nN} \delta (\det \mh - V_0 ) \:  \exp(-\beta V_0 \mt:\meta)  \nonumber \\
 & \times & \prod_{i=1}^n \int \mathcal{D}\XX_i(s) \exp(-\beta H) \prod_{\xx} \delta ( \rhohat_A (\xx ) +\rhohat_B (\xx
 ) -nN) \label{partf}
\end{eqnarray}
($\mt :\meta$ is the contraction
$t_{\alpha\beta}\epsilon\alpha\beta$). The final factor in the
above expression imposes the constraint of local
incompressibility. Moreover, incompressibility implies globally
that the cell volume remains fixed at its initial value, i.e.
$\det \mh = V = V_0 = \det \mh_0$. This is enforced by the delta
function in the first line above. Hence all shape transformations
should   be volume preserving. The practical implementation of
this constraint will be discussed below.

Hubbard-Stratonovich transformations are used to convert the
particle-based partition function \ref{partf} into a field theory
\cite{ghf02}. These can be carried out straightforwardly on the
polymer partition function for a given cell shape $\mh$, $Z(\mh
)$, with the result
\begin{eqnarray}
Z(\mh) & \equiv  &  \prod_{i=1}^n \int \mathcal{D}\XX_i(s)
\exp(-\beta H) \prod_{\xx} \delta ( \rhohat_A (\xx ) +\rhohat_B
(\xx
 ) -nN) \nonumber \\
 & = & \int \mathcal{D}w \; \exp(  n \ln Q[w,\mh]-E[w]) \label{zpol}
 \end{eqnarray}
 where $Q[w,\mh]$ is the partition function of a single copolymer chain experiencing a
 chemical potential field $w (\xx ,s )$, $\int \mathcal{D}w$ denotes a functional
 integral over the field $w$, and $E[w]$ is a local quadratic functional of $w$ that reflects
 the A-B monomer interactions and the local incompressibility constraint: \cite{ghf02}
\begin{equation}\label{eqghf1}
    E[w] = \frac{n}{2} \int d\xx \; \left[ \frac{1}{2 \chi N} (w_B
    -w_A )^2 - (w_A + w_B ) \right]
\end{equation}
Here we have noted that for an AB diblock copolymer melt, the
potential $w(\xx ,s)$ amounts to a two-component potential, i.e.
$w(\xx ,s) =w_A (\xx )$ for $s \in [0,f]$ and $w(\xx ,s) =w_B (\xx
)$ for $s \in [f,1]$.

The object $Q[w,\mh]$ is a normalized partition function for a
single copolymer experiencing a potential field $w(\xx ,s)$  This
partition function  can be obtained from a single-chain propagator
$q (\xx ,s)$ that is the solution of a modified diffusion equation
\begin{equation}
\frac{\partial q}{\partial s} = R_{g0}^2 (\mG^{-1})_{\alpha\beta}
\frac{\partial^2 q}{\partial x_\alpha \partial x_\beta} -w(\xx,s)
q(\xx,s) \label{ghfeq11}
\end{equation}
subject to $q(\xx ,0)=1$. The single chain partition function is
given by $Q[w,\mh ]= \int d\xx \; q(\xx,1)$.

Finally, the partition function for an incompressible diblock
copolymer melt confined to a cell of variable  shape can be
expressed as a field theory in the variables $\mh$ and $w$:
\begin{equation}
 Z  =  \int d(\mh) (\det\mh)^{nN}\int \mathcal{D}w \; \delta (\det \mh - V_0 ) \:
  \exp (-F[w,\mh ])
  \label{ghfeq3}
\end{equation}
where $F[w,\mh]$ is an effective Hamiltonian given by
\begin{equation}\label{ghfeq4}
    F[w,\mh] = \beta V_0 \mt:\meta +E[w] - n \ln Q[w,\mh ]
\end{equation}
 In the
mean-field approximation (SCFT), for a given shape $\mh$ of the
simulation box, we approximate the functional integral over $w$ in
eq \ref{ghfeq3} by the saddle point method. For this purpose, the
functional $Q[w,\mh ]$ can be evaluated for any $w$ and $\mh$ by
solving the modified diffusion equation (using e.g. a
pseudo-spectral approach). The saddle point (mean-field) value of
$w$, $w^*$, is obtained by applying a relaxation algorithm
\cite{ghf02,Sides03} to solve
\begin{equation}\label{ghfeq9}
    \left. \frac{\delta F[w,\mh ]}{\delta w(\xx ,s)}
    \right|_{w=w^*} =0
\end{equation}
In the mean-field approximation, $F[w^* ,\mh ]$ corresponds to the
free energy of the copolymer melt (in units of $k_B T$).

In a simulation at constant tension, the relaxation equation for
the fields must be supplemented by a corresponding evolution
equation for the cell. This equation is chosen to be a simple
relaxation

\def\uD{\underline{D}}
\def\uM{\underline{M}}
\def\uI{\underline{1}}
\begin{equation}\label{evolveh}
\frac{d \mh}{dt} = -\lambda_0 \: \mh  \uD \mh^{-1} \frac{\partial
F[w,\mh ]}{\partial \mh}
\end{equation}
where the tensor $\uD$ is a projection operator whose action on an
arbitrary tensor $\uM$ is a traceless tensor, i.e. $\uD \: \uM
\equiv \uM - (1/3) \mathrm{Tr} (\uM ) \uI$. Equation \ref{evolveh}
corresponds to a cell shape relaxation that (for $\lambda_0 >0$)
is down the gradient $\partial F /\partial \mh$, approaching a
local minimum of the mean-field free energy $F[w^* ,\mh ]$. The
``mobility'' tensor $\mh \uD \mh^{-1}$ is chosen so that the cell
shape dynamics described by eq \ref{evolveh} conserves the cell
volume.

Application of eq \ref{evolveh} requires an expression for the
thermodynamic force $\partial F/\partial \mh$. Explicit
differentiation, noting the constraint of constant $\det \mh$,
leads to
\begin{equation}\label{ghfeq6}
    \frac{\partial F[w,\mh ]}{\partial \mh} = \beta V_0
    \left( \frac{\partial}{\partial \mh}\mathrm{Tr} (\meta \mt)
 + \mh \underline{\Sigma} \right)
\end{equation}
where $\underline{\Sigma}$ is a symmetric tensor defined by
\begin{eqnarray}
   \Sigma_{\alpha\beta} [w,\mh ] &  = & - \frac{2 k_B T n}{V}
   \frac{\partial \ln Q[w,\mh ]}{\partial G_{\alpha \beta}}
   \nonumber \\
   & = & \frac{k_BT n}{2V R_{g0}^2}
   \left< \int_0^1 ds
   \frac{dX_{\alpha} (s)}{ds}\frac{dX_{\beta} (s)}{ds}\right> \label{stress}
\end{eqnarray}
The angular brackets in the second expression denote an average
over all conformations $\XX (s)$ of a single copolymer chain that
is subject to a prescribed chemical potential field $w$ and fixed
cell shape $\mh$.

The first term on the right hand side of eq \ref{ghfeq6} can be
conveniently rewritten as
\begin{equation}
\frac{\partial}{\partial h_{\alpha\beta} }(\mathrm{Tr} (\meta
\mt)) = \frac{\partial}{\partial h_{\alpha\beta} }
(\mathrm{Tr}{1\over 2} \mht_0^{-1} \mht\mh \: \mh_0^{-1} \mt) =
(\mh \: \mh_0^{-1}\mt \mht_0^{-1})_{\alpha\beta}.
\end{equation}
Hence, eq \ref{evolveh} can be compactly expressed
as\begin{equation}\label{evolveh1}
    \frac{d\mh}{dt} = -\lambda \mh \uD \left[(\mh_0^{-1}\mt \mht_0^{-1})
    + \mSigma\right]
\end{equation}
where $\lambda >0$ is a new relaxation parameter defined by
$\lambda = \beta V_0 \lambda_0$.

Equation \ref{evolveh1} will evolve the cell shape to a
configuration of minimum free energy (in the mean-field
approximation). This configuration can either be metastable (local
minimum) or stable (global minimum). Addition of a noise source to
the equation provides a means for overcoming free energy barriers
between metastable and stable states, i.e. a simple simulated
annealing procedure.

 An equilibrium solution of the cell shape
equation \ref{evolveh1} is evidently obtained when
\begin{equation}\label{equil}
(\mh_0^{-1}\mt \mht_0^{-1}) +  \mSigma = 0
\end{equation}
Combining eqs \ref{tsigma}, \ref{stress} and \ref{equil}, it is
seen that this equilibrium condition corresponds to a balance
between the \emph{externally} applied Cauchy stress, $\msig$, and
the \emph{internal} elastic stress, $\msig^{int}$, sustained by
the polymer chains
\begin{equation}\label{ghfeq10}
    \msig + \msig^{int} =0
\end{equation}
where
\begin{equation}\label{smallsig}
\sigma_{\alpha\beta}^{int} [w,\mh] \equiv (\mh \mSigma \mht
)_{\alpha\beta} = \frac{k_B T}{2V R_{g0}^2} \sum_{i=1}^{n}
   \left< \int_0^1 ds \frac{dR_{i\alpha}}{ds}\frac{dR_{i\beta}}{ds} \right> \ .
\end{equation}
This expression for the internal polymer stress is well-known in
the polymer literature \cite{doiedwards}.

Equation \ref{evolveh1} drives a change in the shape of the
simulation cell (at constant cell volume) to approach the
equilibrium condition \ref{equil} at which the internal elastic
stress of the copolymers balances the imposed external stress.

The last step  is to find an expression for the internal stress
tensor $\msig^{int}$ (eq \ref{smallsig}) or $\mSigma$ (eq
\ref{stress}) in terms of the single chain propagator, which is
the central object computed in a field-theoretic simulation
\cite{ghf02}. The appropriate expression turns out to be
\begin{equation}\label{stressfinal2}
   \frac{ \sigma_{\alpha \beta}^{int}}{(n/V) k_B T} = -
   \frac{2 R_{g0}^2}{ Q}
h^{-1}_{\gamma \alpha} \int d\XX \int_0^1 ds \;
 \frac{\partial q(\XX,s)}{\partial X_\gamma} \frac{\partial q(\XX,1-s)}{\partial
 X_\delta} h^{-1}_{\delta\beta}
\end{equation}
we refer to the appendix and to \cite{bfs05} for the derivation of
this expression.

From the above derivations, it is clear that the quantity that is
externally imposed in the method is not the Cauchy stress, but
rather the thermodynamic tension. The Cauchy stress, which is the
experimentally accessible quantity, is a result of the simulation,
as is the cell shape. This feature is general in any application
of the Parrinello-Rahman method in which a partition function of
the form ref{zpol} is sampled using Monte-Carlo, Langevin or
molecular dynamics. The Cauchy stress has therefore to be obtained
independently, using equation \ref{stressfinal} in the present
case, or in the case of molecular system through the
Irving-Kirkwood formula.

\begin{figure}
\begin{center}
\begin{tabular}{ccc}
\hspace{-5cm} \includegraphics[height=12cm,angle=-90]{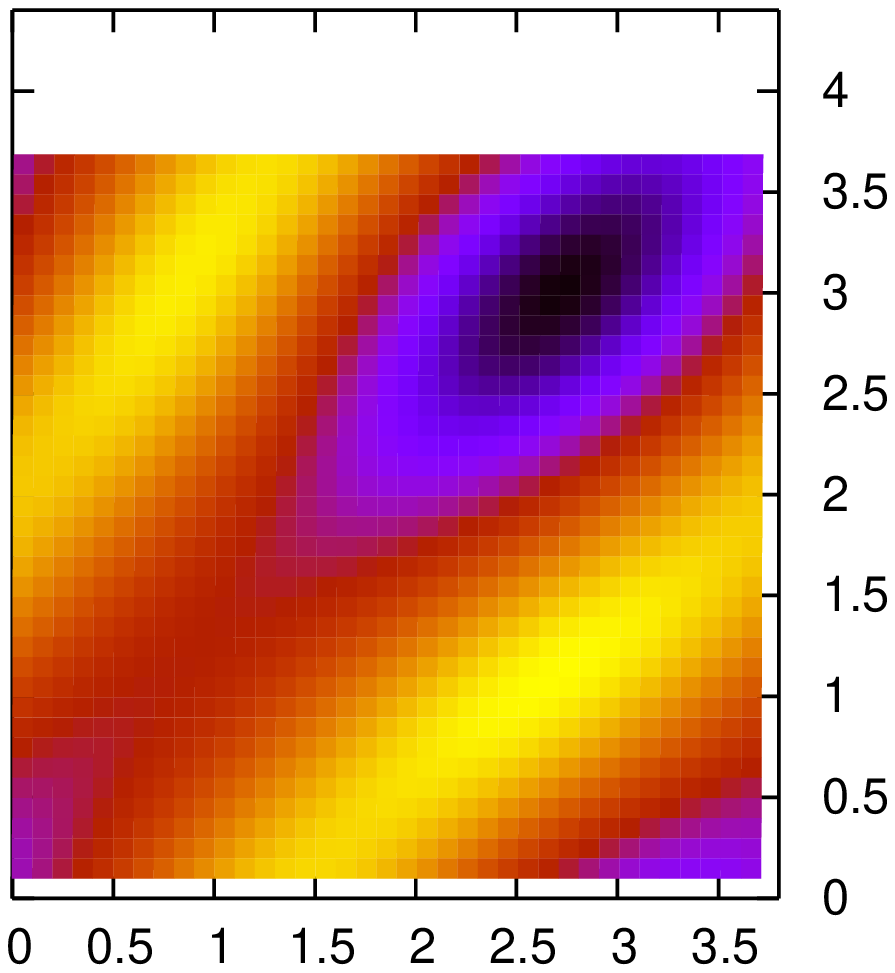}&
\hspace{-8cm}  \includegraphics[height=12cm,angle=-90]{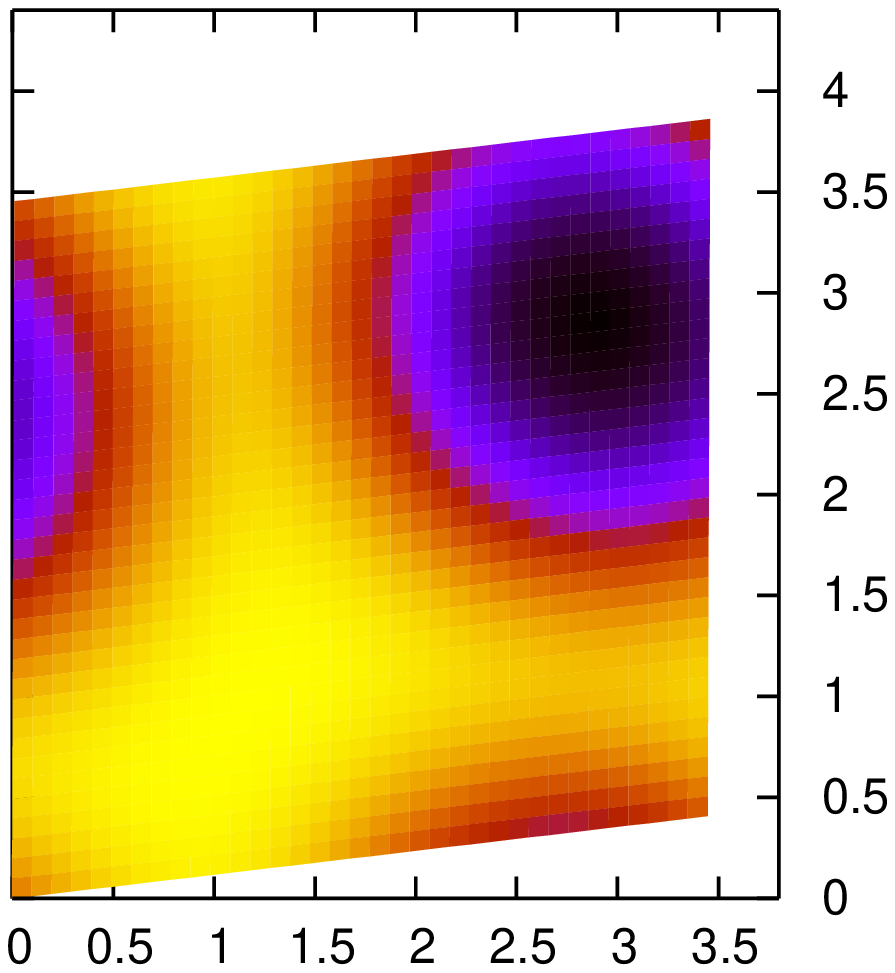}&
\hspace{-8cm} \includegraphics[height=12cm,angle=-90]{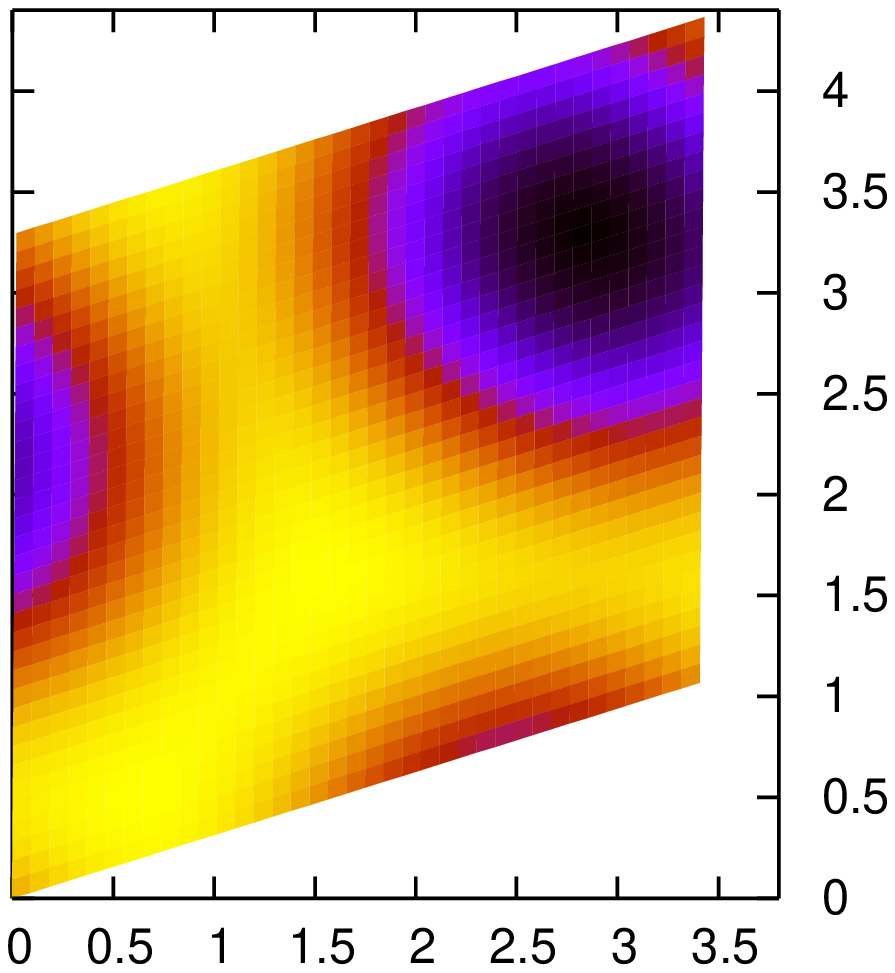}
\\
\end{tabular}
\end{center}
\caption{Transformation of a square cell under zero external
stress, when the melt is quenched into the stability region of the
cylindrical phase ($\chi N=15.9$, $f=0.64$).} \label{fig5}
\end{figure}

To illustrate the method,  figure \ref{fig5} shows  the evolution
of the simulation cell under zero tension in a simple case. In the
system under study, the parameters have been chosen so that the
equilibrium phase is ordered on a triangular lattice, under zero
external stress. Starting with a square simulation cell, evolution
to the correct rhombohedral shape is obtained after a few
relaxation steps. Other examples of application may be found in
\cite{bfs05}.

\section*{Acknowledgments}
\addcontentsline{toc}{section}{Acknowledgements} This work was
supported by the CNRS. The work presented in these notes results
from collaborations with F. Leonforte, A. Tanguy, J. Wittmer, L.J.
Lewis, S.W Sides and G.H. Fredrickson. A very careful reading of
the manuscript by the referee  is also acknowledged. Finally,  I
would like to thank Mike Klein for the many stimulating
discussions we had, over the last 20 years, on various topics,
including  -but not restricted to!- elasticity and  glasses.

\appendix
\section{Expression for the stress tensor in SCFT}

To obtain equation \ref{stressfinal2},  we start with  the
definition of $\mSigma$
\begin{equation}\label{stress1}
 \beta V \Sigma_{\alpha\beta} = - 2 n
  \frac{1}{Q}\frac{\partial Q[w,G]}{\partial G_{\alpha \beta}}
\end{equation} The derivative of the single chain
partition function can be calculated by discretizing the paths
with a small contour step $\Delta$.
\begin{eqnarray}\label{}
 -\frac{2}{Q}\frac{\partial Q[w,G]}{\partial G_{\alpha \beta}}=&
 \frac{1}{2R_{g0}^2 Q} \int_0^1 ds \int d\XX \int d\XX'\cr
& \int \mathcal{D}\XX(s) \delta(\XX-\XX(s))
 \delta(\XX'-\XX(s+\Delta))
 \left(\frac{X_\alpha-X_\alpha'}{\Delta}\right)
 \left(\frac{X_\beta-X_\beta'}{\Delta}\right)\cr
&\exp\left(-\frac{1}{4 R_{g0}^2}\int_0^1 ds
\frac{dX_{\alpha}(s)}{ds}  G_{\alpha\beta}
\frac{dX_{\beta}(s)}{ds} - \int_0^1 ds \: w(\XX(s),s)\right)
 \end{eqnarray}
Except between the points $\XX,s$ and $\XX',s+\Delta$ one can
replace the path integrals with propagators $q$, so that
\begin{eqnarray}\label{}
 -\frac{2}{Q}\frac{\partial Q[w,G]}{\partial G_{\alpha \beta}}=&
\frac{1}{2R_{g0}^2 Q} \int_0^1 ds \int d\XX \int d\XX' q(\XX,s)
q(\XX',1-s-\Delta) \cr
 &\left(\frac{X_\alpha-X_\alpha'}{\Delta}\right)
 \left(\frac{X_\beta-X_\beta'}{\Delta}\right)
 \exp\left( -\frac{1}{4\Delta R_{g0}^2}G_{\alpha\beta}(X_\alpha-X_\alpha')(X_\beta-X_\beta')\right) \cr
 \end{eqnarray}

One can then set $\XX'=\XX+\uu$, and expand for small $\uu$ and
$\Delta$ according to
\begin{equation}\label{}
q(\XX+\uu,1-s-\Delta) = q(\XX,1-s) -\Delta \frac{\partial
q}{\partial s} + u_\gamma \frac{\partial q}{\partial X_\gamma }+
\frac{1}{2}u_\gamma u_\delta \frac{\partial^2 q}{\partial X_\gamma
\partial X_\delta}
\end{equation}
The derivative w.r.t. $s$ can be eliminated by applying the
modified diffusion eq \ref{ghfeq11}.  One also requires second and
fourth moments of the Gaussian distribution of displacements
$\uu$,
 $$\overline{u_\alpha u_\beta} = G^{-1}_{\alpha\beta} (2 R_{g0}^2 \Delta)$$
 $$\overline{u_\alpha u_\beta u_\gamma u_\delta}=
 (2 R_{g0}^2 \Delta)^2 (G^{-1}_{\alpha\beta}G^{-1}_{\gamma\delta}
+G^{-1}_{\alpha\gamma}G^{-1}_{\beta\delta} +
G^{-1}_{\alpha\delta}G^{-1}_{\beta\gamma})$$

By means of these results, we have
\begin{eqnarray}\label{}
 -\frac{2}{Q}\frac{\partial Q[w,G]}{\partial G_{\alpha \beta}}=&
 G^{-1}_{\alpha\beta}\frac{ \int d\XX\rho(\XX)}{\Delta}
 +  G^{-1}_{\alpha\beta} \int d\XX w(\XX) \rho(\XX) \cr
 - &\frac{R_{g0}^2}{Q} G^{-1}_{\alpha\beta} G^{-1}_{\gamma\delta} \int d\XX \int_0^1 ds
 q(\XX,s)\frac{\partial^2 q}{\partial x_\gamma \delta
 x_\beta}\cr
 + &
 \frac{R_{g0}^2}{Q} (G^{-1}_{\alpha\beta}G^{-1}_{\gamma\delta}+
G^{-1}_{\alpha\gamma}G^{-1}_{\beta\delta} +
G^{-1}_{\alpha\delta}G^{-1}_{\beta\gamma})\int d\XX \int_0^1 ds
 q(\XX,s) \frac{\partial^2 q}{\partial x_\gamma \partial
 x_\delta}\cr &
\end{eqnarray}
where $\rho(\XX) =Q^{-1} \int_0^1 ds \: q(\XX ,s)q(\XX,1-s)$ is
the single-chain total monomer density operator. There is a
partial cancellation in the last two terms so that
\begin{eqnarray}\label{}
 -\frac{2}{Q}\frac{\partial Q[w,G]}{\partial G_{\alpha \beta}}=&
 G^{-1}_{\alpha\beta}\frac{ \int d\XX\rho(\XX)}{\Delta}
 + i G^{-1}_{\alpha\beta} \int d\XX w(\XX) \rho(\XX) \cr
 + &
 \frac{2 R_{g0}^2}{Q}
G^{-1}_{\alpha\gamma}G^{-1}_{\beta\delta} \int d\XX \int_0^1 ds
 q(\XX,s) \frac{\partial^2 q}{\partial X_\gamma \partial
 X_\delta}\cr &
\end{eqnarray}

The internal polymer stress is obtained after matrix
multiplication by $\mh$ on the left and $\mht$ on the right. This
implies that the first two terms become a simple isotropic stress
contribution, and are therefore not relevant to an incompressible
system. The final formula for the internal stress tensor is
therefore, apart from this diagonal contribution,
\begin{equation}\label{stressfinal}
    \sigma_{\alpha \beta}^{int} =\left(\frac{nk_BT}{V}\right)\frac{2 R_{g0}^2}{Q}
h^{-1}_{\gamma \alpha} \int d\XX \int_0^1 ds \;
 q(\XX,s) \frac{\partial^2 q(\XX,1-s)}{\partial X_\gamma \partial
 X_\delta} h^{-1}_{\delta\beta}
\end{equation}
The tensor $\Sigma$ appearing in equation  \ref{evolveh1} is given
by an
 expression similar to \ref{stressfinal}, with $\mG$ replacing $\mh$.
 The factor $k_B T n/V$ accounts for the total number of chains,
and produces a stress with the correct dimensions.  In practice,
the stress will be made dimensionless by dividing by this factor,
so that the dimensionless stress is given by
\begin{equation}\label{stressfinal1}
   \frac{ \sigma_{\alpha \beta}^{int}}{(n/V) k_B T} =
   \frac{2 R_{g0}^2}{ Q}
h^{-1}_{\gamma \alpha} \int d\XX \int_0^1 ds \;
 q(\XX,s) \frac{\partial^2 q(\XX,1-s)}{\partial X_\gamma \partial
 X_\delta} h^{-1}_{\delta\beta}
\end{equation}
Equation \ref{stressfinal2} is obtained after integrating by
parts.

 A local
(rather than volume averaged) version of this connection between
the stress tensor and the polymer propagator was derived
previously in \cite{Fredrickson02}. Numerically, $\sigma_{\alpha
\beta}$ is evaluated from eq \ref{stressfinal1} using a
pseudo-spectral scheme. The derivatives with respect to spatial
coordinates are obtained by multiplying the propagator by the
appropriate components of the wavevector in Fourier space, and
transforming back into real space.

\bibliographystyle{unsrt}

\end{document}